\documentclass[graybox, nosecnum]{svmult}

\usepackage{mathptmx}       
\usepackage{helvet}         
\usepackage{courier}        
\usepackage{type1cm}        
\usepackage{makeidx}         
\usepackage{graphicx}        
\usepackage{multicol}        
\usepackage[bottom]{footmisc}
\usepackage[hidelinks,colorlinks=true,citecolor=Blue]{hyperref}        
\usepackage{soul}            
\usepackage{amssymb} 
\usepackage{amsmath} 
\usepackage{braket}
\usepackage{breakcites}
\usepackage[dvipsnames]{xcolor}
\makeindex             

\begin{document}
\title*{Many-body collective neutrino oscillations: recent developments}
\author{Amol V. Patwardhan, Michael J. Cervia, Ermal Rrapaj, Pooja Siwach, A.~B.~Balantekin
}
\institute{Amol V. Patwardhan \at SLAC National Accelerator Laboratory, 2575 Sand Hill Rd, Menlo Park, CA 94025 \\ \email{apatward@slac.stanford.edu}
 \and Michael J.~Cervia \at George Washington University, 725 21st St NW, Washington, DC 20052 \\ \email{cervia@gwu.edu}
 \and Ermal Rrapaj \at University of California, Berkeley, CA 94720-7300 \\ \email{ermalrrapaj@berkeley.edu}
 \and Pooja Siwach \at University of Wisconsin, 1150 University Ave, Madison, WI 53706 \\
 \email{psiwach@physics.wisc.edu}
 \and A.B. Balantekin \at University of Wisconsin, 1150 University Ave, Madison, WI 53706 \\ \email{baha@physics.wisc.edu}
}
%
%
\maketitle
 \abstract{
Neutrino flavor transformations in core-collapse supernovae and binary neutron star mergers represent a complex and unsolved problem that is integral to our understanding of the dynamics and nucleosynthesis in these environments. The high number densities of neutrinos present in these environments can engender various collective effects in neutrino flavor transformations, driven either by neutrino-neutrino coherent scattering, or in some cases, through collisional (incoherent) interactions. An ensemble of neutrinos undergoing coherent scattering among themselves is an interacting quantum many-body system---as such, there is a tantalising prospect of quantum entanglement developing between the neutrinos, which can leave imprints on their flavor evolution histories. Here, we seek to summarize recent progress that has been made towards understanding this phenomenon.
}


\section{Motivation: Supernovae, Mergers, and the Early Universe}

In extreme astrophysical environments such as core-collapse supernova explosions, and binary neutron star (or black hole - neutron star) mergers, as well as during certain epochs in the early universe, neutrinos dominate the transport of energy, entropy, and lepton number (for example, see ~\cite{Janka:2006fh, Burrows:2020qrp, Fuller:2022nbn, Foucart:2022bth, Kyutoku:2017voj, Grohs:2015tfy}, etc.). The key processes governing neutrino transport in these environments are electron neutrino and antineutrino captures on nucleons, i.e.,
\begin{align}
    \nu_e + n \rightleftharpoons p + e^- \\
    \bar\nu_e + p \rightleftharpoons n + e^+
\end{align}
A consequence of the typical temperatures and densities of these environments is that neutrinos decouple with energies of $\mathcal{O}(1\text{--}10)\,$MeV, and therefore, the $\mu$ and $\tau$ flavor (anti-)neutrinos are unable to participate in these charged-current processes, due to there not being enough energy to produce $\mu$ and $\tau$ leptons in the final state. Given the importance of these processes in the energy transport, as well as in determining the neutron-to-proton ratio and the resulting nucleosynthesis prospects (e.g., ~\cite{Surman:2003qt,Martinez-Pinedo:2017ksl,Kajino:2014bra,Frohlich:2015spx,Langanke:2019ggn,Roberts:2016igt,Steigman:2012ve,Grohs:2015tfy}), the flavor-asymmetric nature of charged-current capture necessitates a thorough understanding of neutrino flavor evolution in these environments. The potential impact of neutrino flavor evolution on nucleosynthesis has already been studied in various contexts (e.g., \cite{Qian:1993dg,Yoshida:2006qz,Duan:2010af,Kajino:2012zz,Wu:2014kaa,Sasaki:2017jry,Balantekin:2017bau,Xiong:2019nvw,Xiong:2020ntn}).

In what follows, we shall summarize recent progress in our understanding of a particular facet of neutrino oscillations in extreme astrophysical environments---namely, the quantum many-body nature of collective neutrino oscillations engendered by $\nu$-$\nu$ interactions in dense neutrino streams.

\section{Introduction to collective neutrino oscillations}

The neutral current weak term of the Standard Model (SM) allows neutrinos to interact pairwise via virtual $Z$-boson exchange or, more simply, in the low-energy effective theory, via the Fermi four-point interaction
\begin{equation} \label{eq:nunuint}
    H_\mathrm{int} \equiv \frac{G_F}{\sqrt{2}} \sum_{f,g}\overline{\nu}_g\gamma^\mu\nu_g\overline{\nu}_f\gamma_\mu\nu_f,
\end{equation}
where $f,g$ span the flavor state indices. The relevance of these interactions in environments where the number densities of neutrinos are comparable to (or larger than) those of charged leptons, e.g., in core-collapse supernovae, binary neutron star mergers, as well as in the early universe, had been discussed in~\cite{Notzold:1987ik,Fuller:!987aa}. But the extent of their importance in changing the flavor content of neutrinos, via diagonal \textit{and off-diagonal} contributions to the neutrino Hamiltonian, was not fully recognized until later~\cite{Pantaleone:1992xh,Pantaleone:1992eq,Samuel:1993}. 

Considering pairs of neutrinos with well-defined incoming momenta $\mathbf{p}$ and $\mathbf{q}$ (i.e., plane wave states) and the same pair of outgoing momenta (i.e., \lq\lq forward scattering\rq\rq\ neutrinos, the contributions of which can be added coherently), the off-diagonal matrix elements of the interaction Hamiltonian $H_\mathrm{int}$ may be interpreted as arising from \lq\lq flavor swaps\rq\rq\ between neutrino pairs (in the flavor basis). Because the off-diagonal term exchanges flavor between the \lq\lq test\rq\rq\ and the \lq\lq background\rq\rq\ neutrinos, the flavor evolution of the interacting neutrinos constitutes a many-body problem, potentially rendering the one-particle propagation formalism~\cite{Samuel:1993,Sigl:1993ctk,Qian:1994wh} inadequate for describing the resulting dynamics.
Notably, the interaction Hamiltonian $H_\text{int}$ does not commute with the Hamiltonian terms corresponding to flavor oscillations in vacuum and neutrino interactions with background matter.  Consequently, in a regime where the strength of these terms is comparable in scale to the $\nu$-$\nu$ interaction strength, 
diagonalizing this Hamiltonian is not straightforward and the many-body problem acquires a nontrivial nature. 
Here, the entire Hilbert space of $N$ interacting neutrinos and antineutrinos in $n_f$ flavors has dimension $n_f^N$. 

Despite emphasis on the high nonlinearity of this problem,~\cite{Samuel:1993} had proposed that a statistical mechanical approach, whereby a two-flavor neutrino density matrix is treated as interacting with a background of neutrinos and antineutrinos, could describe the evolution of a dense neutrino gas for certain portions of this parameter space. 
This analysis was extended by
\cite{Sigl:1993ctk} to $n_f\geq2$ flavors with proposed evolution of $n_f\times n_f$ density matrices via quantum Boltzmann equations, including collision integrals as well as more general, potentially non-SM coupling between flavors. {In these treatments, the collisional contributions can lead to a nontrivial loss of coherence being reflected in the density matrices of individual neutrinos.
However, the ability to calculate multi-body wave functions that exhibit $\nu$-$\nu$ correlations is relinquished, in exchange for a more favourable scaling of computational complexity with the number of neutrinos in the simulation.} Along these lines,~\cite{Qian:1994wh} proposed a physical ansatz that the wave function of the ensemble is not a coherent many-body state, but simply composed of single-neutrino wave functions with random relative phases, to be summed incoherently, called the Random Phase Approximation (RPA). In this way, each neutrino density matrix is taken to be pure, and the effective Hilbert space dimension is reduced to $n_fN$. In kind, the complexity of collective oscillations calculations becomes greatly simplified. {This ansatz amounts to a \lq\lq mean field approximation\rq\rq\ wherein expectation values of operator products may be replaced by products of the individual operator expectation values.}
Notably, 
this physical description of neutrinos expressly prohibits the quantum entanglement between neutrinos. 
As such, assessing the validity of this ansatz involves determining the extent to which quantum effects are needed to correct this approximation. In this chapter, we discuss recent progress along this front.

Before delving into the chapter, we mention in passing that recent years have seen a rapid growth of interest in flavor instabilities and resulting fast flavor oscillation, even within the scope of the mean field approximation. For more information we refer the reader to the chapter on \lq\lq Fast Flavor Transformations\rq\rq\ by~\cite{Richers:2022zug}, or the review articles by~\cite{Chakraborty:2016yeg,Tamborra:2020cul}.

\section{Many-body Hamiltonian for interacting neutrinos}

The Hamiltonian describing a system of interacting neutrinos can be written in terms of generators of $SU(n_f)$, and it possesses a $SU(n_f)$ rotation symmetry in neutrino flavor. A significant feature of $\nu$-$\nu$ interactions is the dependence of the interaction strength on the intersection angle between their trajectories. This dependence introduces a geometric complexity to the problem, in addition to the complexity associated with the exponential scaling of the Hilbert space. 

For simplicity, if we consider neutrino mixing between only two flavors, $\nu_e$ and $\nu_x$, then a Hamiltonian consisting of terms that represent vacuum mixing as well as $\nu$-$\nu$ interactions can be written as
\begin{equation} \label{eq:ham}
H = \sum_{\mathbf{p}}\omega_\mathbf{p} \, \vec{B}\cdot\vec{J}_{\mathbf{p}} + \frac{\sqrt{2}G_{F}}{V}\sum_{\mathbf{p},\mathbf{q}}\left(1-\mathbf{\widehat p}\cdot\mathbf{\widehat q}\right)\,\vec{J}_\mathbf{p}\cdot\vec{J}_\mathbf{q}~,
\end{equation}
where $\vec{B}=(0,0,-1)$ in the mass-basis representation, and $\omega_\mathbf{p} = {\delta m^2}/{(2|\mathbf{p}|)}$ are the vacuum oscillation frequencies for neutrinos with momenta $\mathbf{p}$.  Here, $\mathbf{\widehat p}$ and $\mathbf{\widehat q}$ are the unit vectors along the momenta of the interacting neutrinos, and $V$ is the quantization volume. For ease of notation, one can define a $\nu$-$\nu$ coupling parameter $\mu \equiv \sqrt{2} G_F N/V$, where $N$ is the total number of interacting neutrinos. The operators $\vec{J}_{\mathbf{p}}$ represent the neutrino \lq\lq isospin\rq\rq\ in the mass basis, where isospin up and down correspond to the mass basis states $\ket{\nu_1}$ and $\ket{\nu_2}$.  In this depiction, $\vec{B}$ can be interpreted as a \lq\lq background field\rq\rq\ with which the neutrino isospins interact. Here, we exclude the term representing neutrino interactions with ordinary matter (e.g., charged leptons), since it has a structure that is conceptually similar to the vacuum oscillation term---i.e., consisting of individual neutrinos interacting with a background. In contrast, the $\nu$-$\nu$ interaction term consists of pairs of neutrino isospins interacting with each other. 

In terms of the Fermionic creation and annihilation operators, the neutrino isospins are described as~\cite{Balantekin_2006}
\begin{gather}
	{J}_{\mathbf{p}}^{+}= a_{1}^{\dagger}(\mathbf{p})\,a_{2}(\mathbf{p})~, \qquad
	{J}_{\mathbf{p}}^z=\frac{1}{2}\left(a_{1}^{\dagger}\,(\mathbf{p})a_{1}(\mathbf{p})-a_{2}^{\dagger}\,(\mathbf{p})a_{2}(\mathbf{p})\right)~, 
\end{gather} 
with ${J}_{\mathbf{p}}^{-} = ({J}_{\mathbf{p}}^{+})^\dagger$. In the spin-$1/2$ representation, one can write the isospin operators in terms of Pauli matrices: i.e., $\vec{J}_\mathbf{p} = \vec \sigma_\mathbf{p}/2$, where $\sigma_\mathbf{p}$ is a vector of Pauli matrices defined in the subspace of the neutrino with momentum $\mathbf{p}$.



\section{Path integral formulation}

An assessment of quantum corrections to a mean field picture can in principle be performed via a coherent state analysis, as formulated by~\cite{Balantekin_2006}. 
Schematically, in this procedure, one seeks to calculate the matrix elements of the time evolution operator $U(t_f;t_i)$ for a single neutrino in the basis of $SU(n_f)$ coherent states $\ket{z}$ for neutrinos (and/or antineutrinos) in $n_f$ flavors, equivalent to a path integral  
\begin{align}
    \braket{z_f|U(t_f;t_i)|z_i} = \int \mathcal{D}[z,z^*]\:\exp(\mathrm{i}S[z,z^*])
\end{align}
of the derived action
\begin{align}
    S[z,z^*] = \int_{t_i}^{t_f}\mathrm{d}t\:\left[\braket{ z(t)|(\mathrm{i}\partial_t-H)|z(t)} -\mathrm{i}\log\braket{z_f|z_i} \right],
\end{align}
where $H$ is the Hamiltonian of the many-body system. 
A saddle-point approximation of the resulting path integral yields the classical action, which is in complete agreement with the RPA used to derive the mean field theory for collective neutrino oscillations. 
However, in this perspective, analyzing quantum corrections to this approximation entails a careful analysis of the Hessian matrix of the action integral derived from this procedure. Such mathematical analysis has not yet been presented to date. 



\section{Beyond the Mean-Field: Entanglement, Correlations, and Dynamical Phase Transitions}

\subsection{Early literature}

The seminal work describing the $\nu$-$\nu$ interaction Hamiltonian from Eq.~\eqref{eq:nunuint} recognized that these interactions give rise to a quantum many-body problem, which may not in the general case be factorizable in terms of a one-particle effective approximation~\cite{Pantaleone:1992xh,Pantaleone:1992eq}. Subsequently, there were some attempts to ascertain the validity of the one-particle effective approximation~\cite{Bell:2003mg,Friedland:2003dv,Friedland:2003eh,Friedland:2006ke}. In these works, the flavor evolution of interacting neutrinos was analyzed with two different approaches: (i) using two intersecting beams of neutrinos, where the flavor evolution was described in terms of a sequence of elementary scattering amplitudes, and (ii) using a neutrino ensemble represented as interacting plane waves in a box. 

Following initial disagreement regarding whether substantial quantum entanglement can develop among interacting neutrinos~\cite{Bell:2003mg,Friedland:2003dv}, it was subsequently concluded that the build-up of entanglement and resulting flavor conversion would occur on timescales whose scaling is suggestive of incoherent effects~\cite{Friedland:2003eh}. These conclusions were further generalized in ~\cite{Friedland:2006ke}. However, these analyses nevertheless involved several simplifications, most notably, the omission of the one-body terms in the Hamiltonian. The interplay between vacuum oscillations and $\nu$-$\nu$ interaction terms has been shown to give rise to interesting collective phenomena such as \lq\lq spectral splits\rq\rq~\cite{Duan:2006jv,Duan:2006c,Duan:2007,Raffelt:2007,Raffelt:2007xt}, even in the mean-field approximation. Therefore, studying the quantum many-body dynamics of collective neutrino oscillations, with both one- and two-body terms fully incorporated, remains an interesting problem.

With these seemingly conflicting results in the past predicting either a vanishingly small contribution in the large system size limit~\cite{Friedland:2003eh,Friedland:2006ke} or substantial flavor evolution over time scales $\tau_F \sim \mu^{-1}\log(N)$ that can remain relevant for large systems~\cite{Bell:2003mg,Sawyer:2004}, the role of entanglement and quantum effects in the out-of-equilibrium dynamics~\cite{Eisert:2015} of neutrinos has received renewed interest recently (e.g., \cite{Cervia:2019,Rrapaj:2020} and subsequent works mentioned later in this chapter). Note that flavor oscillations on the time scale $\tau_F$ can be considered to be \lq\lq fast\rq\rq, different from \lq\lq slow\rq\rq\ oscillations occurring over $\tau_L \sim \mu^{-1}\sqrt{N}$. In the literature on collective flavor effects in the mean field approximation, one can more commonly find \lq\lq fast\rq\rq\ and \lq\lq slow\rq\rq\ oscillations associated with time scales $\sim\mu^{-1}$ and $\sim\sqrt{\mu\omega}$ (or $\omega$), respectively. 

\subsection{Single-angle approximation, invariants, and integrability}

To circumvent the geometric complexity of the problem, the frequently-employed \emph{single-angle} approximation replaces the angle-dependent (i.e., $\widehat{\mathbf{p}},\widehat{\mathbf{q}}$-dependent) interaction strengths among pairs of neutrinos with a single, appropriately chosen classical average over the various neutrino trajectories. In this limit, one can define a trajectory-averaged interaction parameter $\mu \equiv (\sqrt{2} G_F N/V) \langle 1 - \mathbf{\widehat p} \cdot \mathbf{\widehat q} \rangle$, and approximate the Hamiltonian as
\begin{equation} \label{eq:saham}
H = \sum_{\omega_\mathbf{p}} \omega_\mathbf{p} \, \vec{B}\cdot\vec{J}_{\omega_\mathbf{p}} + \frac{\mu}{N} \, \vec{J} \cdot \vec{J}~,
\end{equation}
where $\vec{J} = \sum_{\omega_\mathbf{p}} \vec J_{\omega_\mathbf{p}}$ is the total neutrino isospin. Note that, in this limit, the neutrino flavor state becomes trajectory-independent, introducing a considerable simplification in the problem. As a result, the neutrinos may be indexed simply by the magnitudes of their momenta (or equivalently, by their vacuum oscillation frequencies $\omega_\mathbf{p}$), rather than by the momenta themselves (magnitude and direction). 
The $\nu$-$\nu$ coupling in general will depend on time. In the context of supernovae, a commonly employed expression for $\mu$ is derived from the spherically symmetric single-angle neutrino bulb model, first described in \cite{Duan:2006b}:
\begin{equation}
    \mu(r) = \mu_0\left[1-\sqrt{1-\left(\frac{R_\nu}{r}\right)^2}\right]^2,
    \label{eq:samu}
\end{equation}
where $r$ is the distance from the center of a \lq\lq neutrino-sphere\rq\rq\ of radius $R_\nu$, which represents a sharp surface where neutrinos decouple from nuclear matter and begin free streaming outwards from the proto-neutron star. We also define $\mu_0\equiv (G_F/\sqrt{2})(N/V)=\mu(R_\nu)$ to be the interaction strength at the neutrino-sphere. Here, the neutrino emission is assumed to be time-invariant over the short time scales associated with neutrino propagation through the supernova envelope, so the interaction strength depends explicitly only on position, rather than time. In the neutrino-driven wind phase of core-collapse supernovae, which occurs over a time window of $\mathcal{O}(1\text{--}10)$\,s after core bounce, one may expect $R_\nu\simeq 20\,\mathrm{km}$ and $\mu_0\sim10^5\omega_0$, where $\omega_0\sim10^{-16}\,\mathrm{MeV}$ is the scale of the vacuum oscillations. During the shock breakout or \lq\lq neutronization burst\rq\rq\ phase that occurs earlier, around $10$\,ms after core bounce, the proto-neutron star can be more extended, with $R_\nu \gtrsim 50\text{--}60$\,km, but the neutrino luminosity is also much higher, resulting in $\mu_0\sim 10^6\omega_0$.

It has been shown that a single-angle Hamiltonian describing neutrino mixing in vacuum and $\nu$-$\nu$ interactions possesses a number of conserved charges which commute with the Hamiltonian~\cite{Pehlivan:2011}. These are analogous to the \lq\lq Gaudin magnets\rq\rq~\cite{Gaudin76} that had been previously identified as the conserved charges of the pairing-force Hamiltonian in nuclear and condensed-matter physics~\cite{Richardson63, Richardson64, Richardson65}. These conserved charges are related to the \emph{integrability} of the Hamiltonian---meaning that it is possible to obtain, in principle, exact eigenvalues and eigenstates of this Hamiltonian in terms of closed-form solutions to a set of algebraic \lq\lq\emph{Bethe-Ansatz}\rq\rq\ equations~\cite{Bethe31}. Based on these ideas, specific procedures for the eigen-decomposition of a single-angle neutrino Hamiltonian have been outlined in the literature~\cite{Pehlivan:2011, Birol:2018qhx, Patwardhan:2019zta}.

Besides descriptions in terms of instantaneously conserved charges, analogies with other many-body problems have been fruitful to yield an explanation of the neutrino flavor spectral split in terms of a Bardeen-Cooper-Schrieffer (BCS)-Bose-Einstein Condensate (BEC) crossover-like phenomenon~\cite{Pehlivan:2016lxx}, as well as to help provide many-body predictions of a spectral split~\cite{Birol:2018qhx} specifically in the case of an initial many-body wave function with all neutrinos in the electron flavor state.

\subsection{Instabilities and dynamical phase transitions}

Collective neutrino oscillations are generally assumed to be caused by unstable modes in the mean field dynamics generated by the Hamiltonian described in Eq.~\eqref{eq:ham} (for two flavors).
 These instabilities are able to amplify initially small flavor perturbations exponentially fast (e.g., ~\cite{Sawyer:2004,Sawyer:2005jk,Duan:2010,Chakraborty:2016yeg,Izaguirre:2017,Tamborra:2020cul,Richers:2022zug} and references therein). 
The presence of the forward-scattering interaction can allow collective effects to develop when $\mu \gtrsim \omega_\mathbf{p}$, giving rise to interesting phenomena like synchronization~\cite{Pastor:2002,Fuller:2006,Raffelt:2010,AKHMEDOV:2016}, bipolar oscillations ~\cite{Kostelecky:1995,Duan:2006b,Duan:2007b} and spectral splits/swaps~\cite{Duan:2006c,Duan:2007,Raffelt:2007,Dasgupta:2009,Martin:2020}.
 
\begin{figure}[htb]
    \centering
    \includegraphics[width=0.49\textwidth]{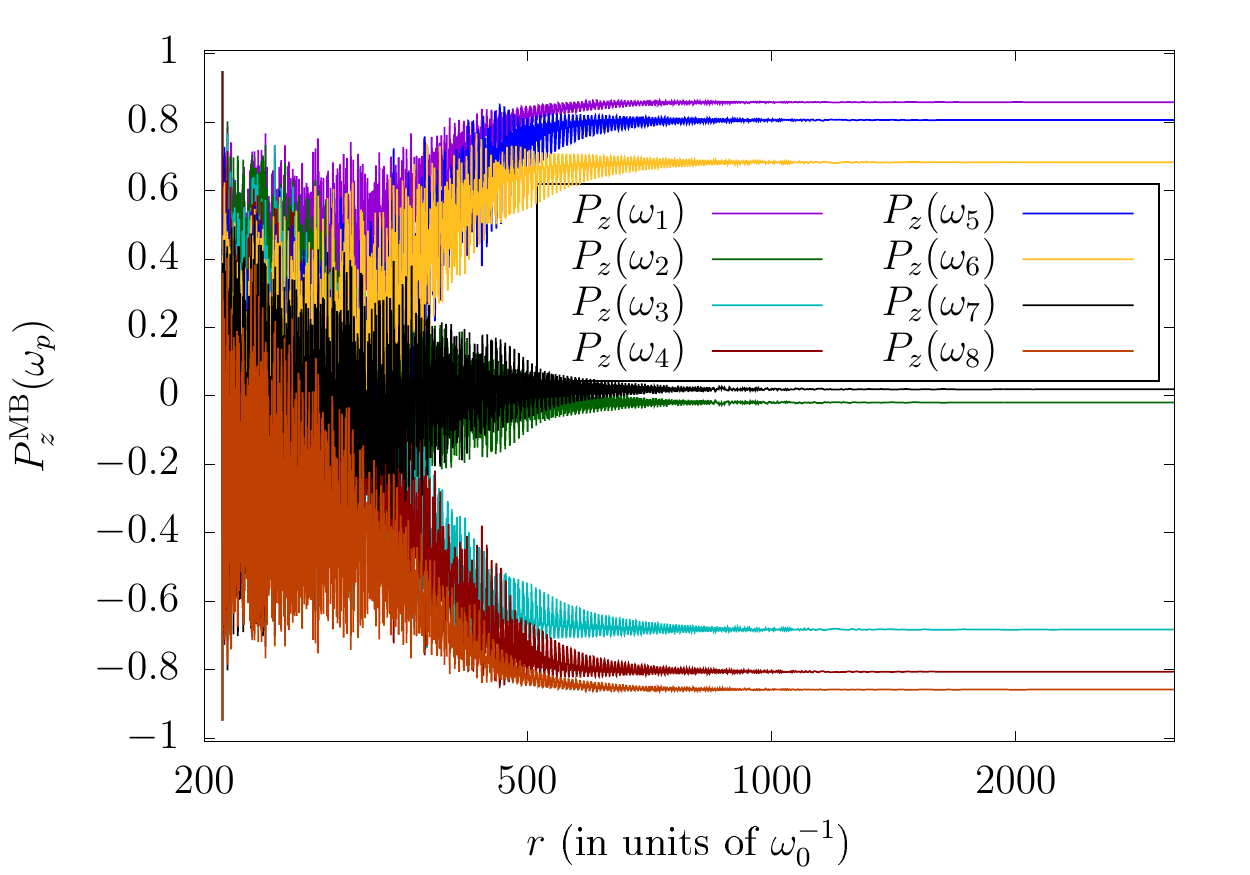}~
    \includegraphics[width=0.49\textwidth]{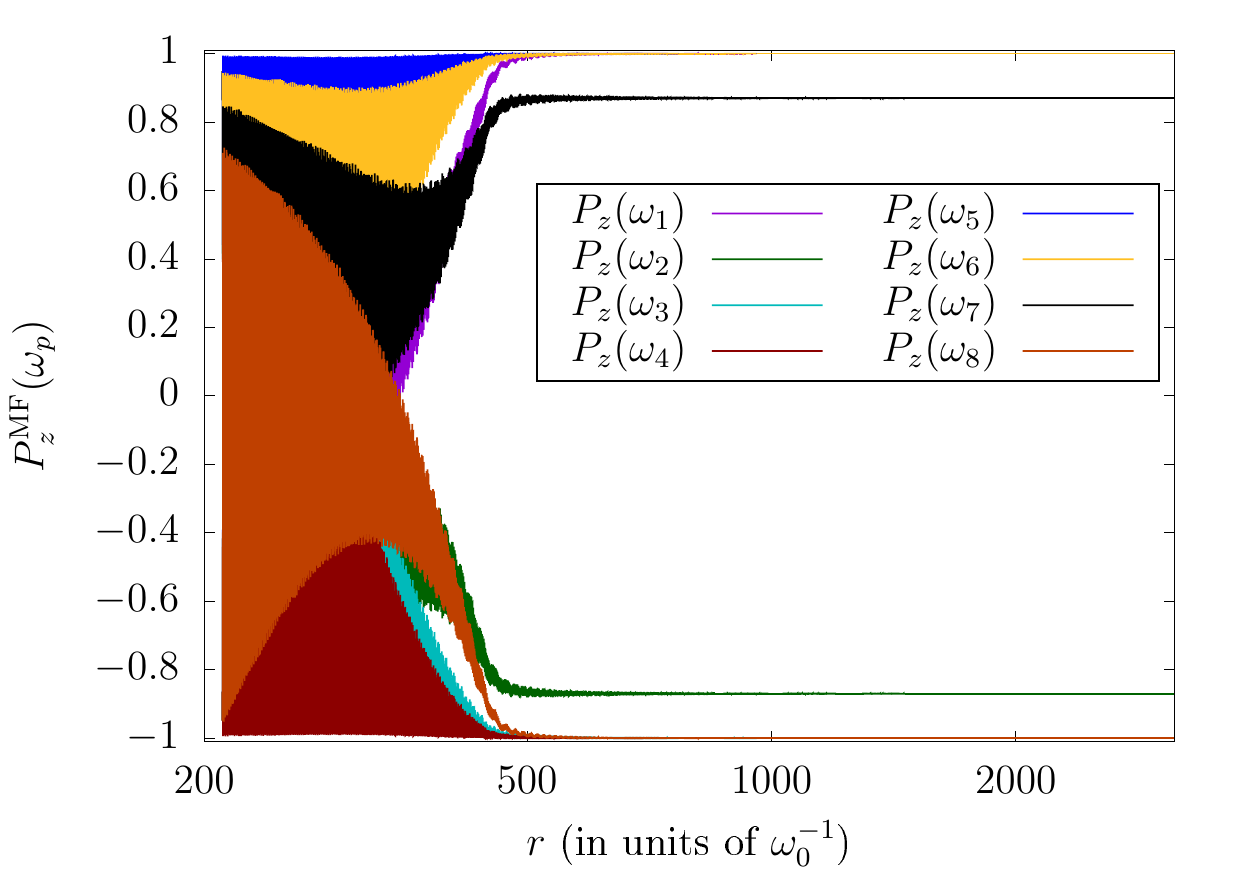}
    \includegraphics[width=0.49\textwidth]{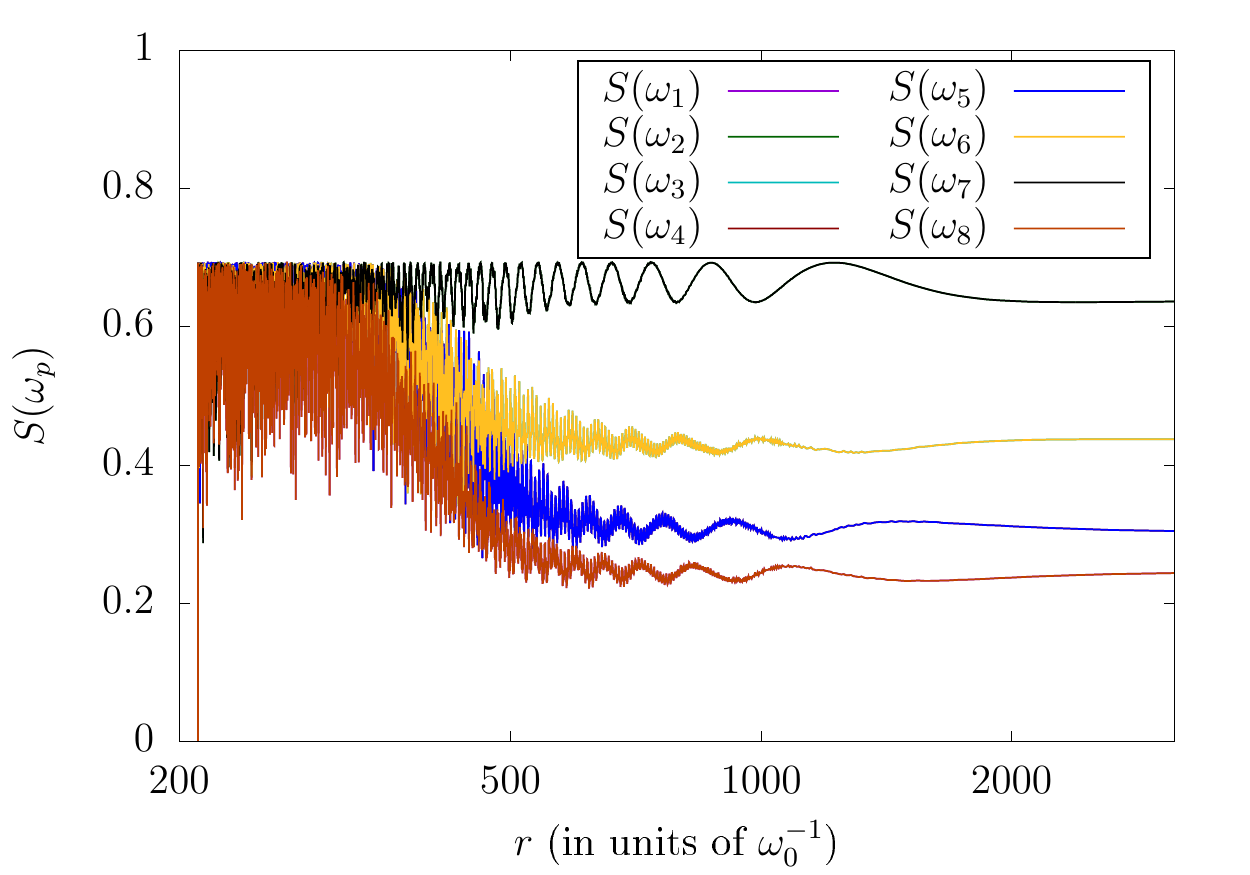}~
    \includegraphics[width=0.49\textwidth]{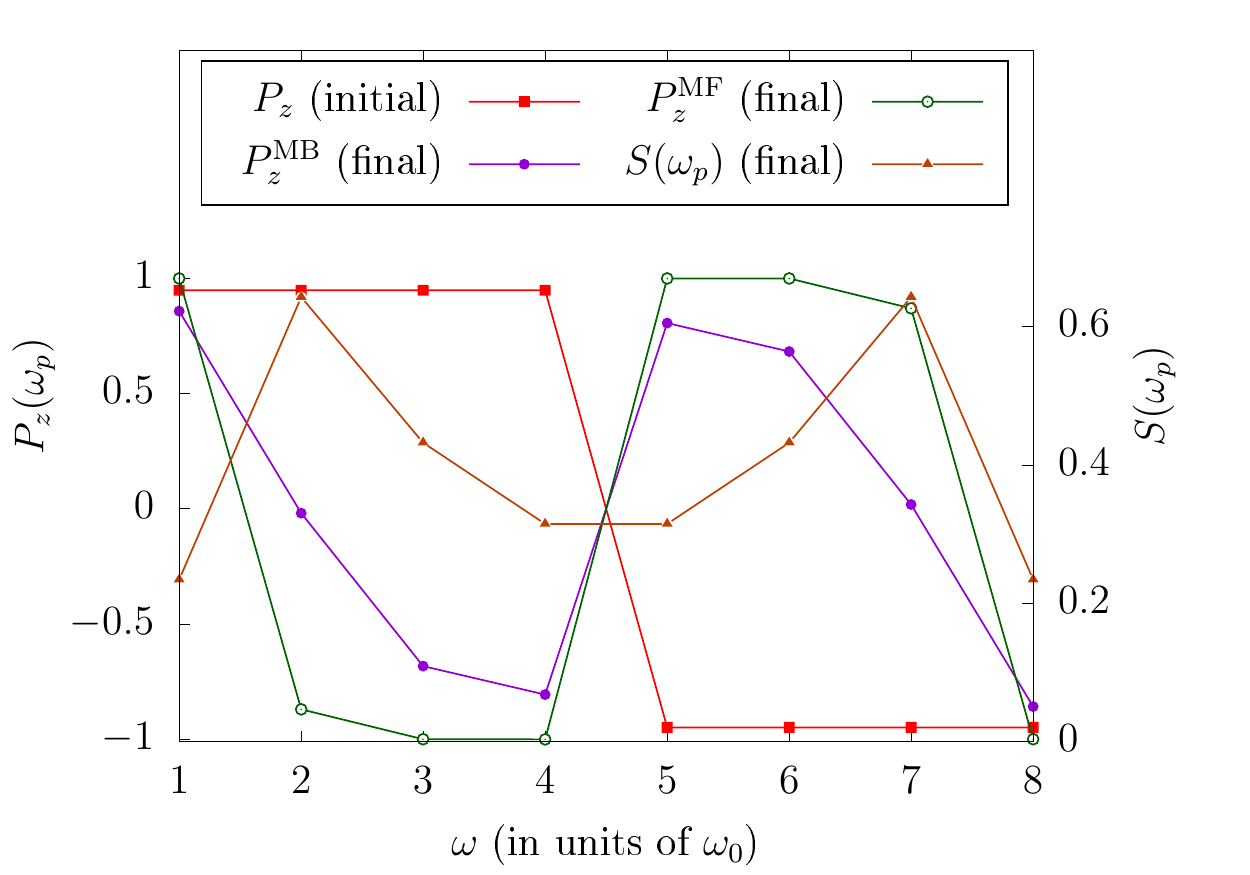}
    \caption{Evolution of an initial state 
    $\ket{\nu_e}^{\otimes4}\ket{\nu_x}^{\otimes4}$ 
    from a starting radius $r_0$ such that $\mu(r_0)=5\omega_0$, 
    with a small mixing angle ($\theta=0.161$) and 
    discrete, equally spaced oscillation frequencies $\omega_k = k\,\omega_0$, and a time-varying neutrino interaction strength $\mu(r)$ motivated by the neutrino bulb model~\cite{Duan:2006c}, in the single-angle approximation according to Eqs.~\eqref{eq:saham} and \eqref{eq:samu}. 
    Details of this calculation can be found in~\cite{Cervia:2019}. 
    \textbf{Top left:} Evolution of the $z$-components of the neutrino isospin expectation values (also known as \lq\lq Polarization vectors\rq\rq) in the mass basis, i.e., $P_z \equiv 2\langle J_z \rangle$, for the full many-body quantum system. \textbf{Top right:} Same as top left, but in the mean-field approximation. \textbf{Bottom left:} Evolution of the entanglement entropy of each neutrino, with respect to the rest of the ensemble. \textbf{Bottom right:} Asymptotic values of $P_z$ vs $\omega_k$, in the full many-body calculation (purple), and in the mean-field approximation (green), together with the initial $P_z$ values (red), and the asymptotic entanglement entropies (dark orange). Neutrinos located closest to the spectral splits in the energy distributions (in this case, at $\omega_2$ and $\omega_7$) develop the largest amount of entanglement and thereby experience the most significant deviations compared to their mean-field evolution.}
    \label{fig8nu}
\end{figure}

On the other hand, in descriptions of interacting neutrino systems that permit many-body quantum dynamics, oscillations that develop on \lq\lq fast\rq\rq\ timescales are generally associated with rapid dynamical development of the neutrino entanglement entropy~\cite{Cervia:2019,Rrapaj:2020,Roggero:2021,Roggero:2021b,Patwardhan:2021rej}. The dynamically generated entanglement between neutrinos is seen to be correlated with deviations from the mean-field dynamics of the system~\cite{Cervia:2019,Rrapaj:2020} and with the presence of spectral splits in the neutrino energy distributions~\cite{Patwardhan:2021rej}. An example of such a calculation is depicted in Fig.~\ref{fig8nu}. In~\cite{Roggero:2022}, rapid entanglement and mean field instabilities were also found to be linked for certain angular setups.

As shown in~\cite{Roggero:2021,Roggero:2021b} in the single angle approximation, when the frequency difference between two neutrino beams ($\delta \omega$) is positive and comparable to the  $\nu$-$\nu$ interaction coupling ($\mu$), $0<\delta \omega \lesssim \mu$, rapid and strong flavor oscillations develop. This rather particular finding can be understood in terms of the presence of a Dynamic Phase Transition (DPT)~\cite{Heyl:2013,Heyl:2018}, which can be characterized by the introduction of the Loschmidt echo,
 \begin{equation}
\label{eq:loch}
\mathcal{L}(t) = \left|\langle \Phi\lvert \exp\left(-\mathrm{i} t H\right)\rvert\Phi\rangle \right|^2\;,
\end{equation}
with $|\Phi\rangle$ the initial state at $t=0$. The quantity $\mathcal{L}(t)$ is a fidelity measure~\cite{Gorin:2006} that quantifies the probability for the system to return to its initial state. A DPT is then characterized by non-analyticities in the rate function
\begin{equation}
\label{eq:loch_rate}
\lambda(t) = -\frac{1}{N}\log\left[\mathcal{L}(t)\right]\;,
\end{equation}
where $N$ is the total number of particles in the system and $\lambda(t)$ an intensive \lq\lq free energy\rq\rq~\cite{Heyl:2013,Gambassi:2012}. Here, the rate $\lambda(t)$ plays the role of a non-equilibrium equivalent of the thermodynamic free-energy. Notably, other definitions of DPT are possible, for instance, time-averaged order parameters~\cite{Sciolla:2011,Sciolla:2013,Zunkovic:2018}. 

\subsection{Phase-space analysis}

In a recent work~\cite{Lacroix:2022krq}, this problem was further explored by analyzing the evolution of neutrino flavor and entanglement in phase space. The setup consisted of two sets (beams) of neutrinos interacting with each other. In this analysis, the Husimi quasi-probability or \lq\lq Q\rq\rq\ representation~\cite{Husimi:1940264} was constructed for the reduced density operator of neutrinos in one of the beams,
using an over-complete basis of coherent states.
In the limit of infinite neutrino number, the Q representation acquires the interpretation of a classical phase-space probability distribution.

For this two-beam interacting neutrino system, it was demonstrated that, while at early times the quasi-probability distribution remains relatively localized, at late times it develops a multi-modal structure with several localized peaks. This delocalization is indicative of non-Gaussian entanglement, which suggests that any approximate method beyond the mean-field relying on only the first and second moments of neutrino observables may not be sufficient to describe the long-term evolution of this system. Based on the phase space analysis, a new method for approximating the exact evolution of the interacting neutrino system was proposed, wherein 
the quantum mechanical many-body evolution is replaced by a statistical average of \lq mean-field\rq\ solutions, with a Gaussian distribution of initial conditions around the exact starting point of the system~\cite{Lacroix:2014sxa}. 

\section{Compact Representations for studying many body effects}

Still allowing for possibilities of mixed one-neutrino density matrices, one proposal~\cite{Volpe:2013uxl} to determine quantum corrections 
is to systematically incorporate $n$-body density matrices $\rho_{1\ldots n}$ for $n \geq 1$, given by
\begin{equation}
    \rho_{1\ldots n} = \frac{N!}{(N-n)!}\mathrm{Tr}_{n+1\ldots N}\rho_{1\ldots N},
\end{equation}
into the coupled equations of motion for $N$ neutrinos, as follows:
\begin{equation}
    \mathrm{i}\partial_t\rho_{1\ldots n} = [H_{1\ldots n},\rho_{1\ldots n}]+\sum_{s=1}^n\mathrm{Tr}_{n+1}[V(s,n+1),\rho_{1\ldots n+1}],
\end{equation}
where $H_{1\ldots n}$ is the Hamiltonian truncated for the first $n$ neutrinos in a given ordering and $V(i,j)$ is the two-body interaction potential for a pair of neutrinos $(i,j)$. 
This procedure is based on the Bogoliubov-Born-Green-Kirkwood-Yvon (BBGKY) hierarchy for density matrices. 
Here, the mean field theory interaction of neutrinos and antineutrinos with the background gas is reproduced 
by restricting to $n=2$ and estimating $\rho_{12}\approx\rho_1\rho_2$ 
(i.e., requiring the two-body correlation function to be zero) 
in this picture, in a sense as a loop Feynman diagram for neutrino propagation. 
In principle, investigating the importance of quantum corrections would practically entail checking for convergence of results for physical observables as the $n$-body correlation functions are incorporated for progressively increasing values of $n$ in the BBGKY hierarchy. 

Owing to the exponential growth in the Hilbert space, classical (conventional) computers are unable to exactly simulate systems of more than $\simeq 20$ neutrinos. To overcome this difficulty, one can resort to compact representations of the wave-function through tensor network methods~\cite{Roggero:2021, Roggero:2021b, Cervia:2022}, and more specifically matrix product states~\cite{Vidal:2003, SCHOLLWOCK:2011, Paeckel:2019}. In simplified setups, these methods allow for the computation of systems of hundreds of neutrinos. Alternatively, when considering very dense neutrino gases (vacuum oscillations can be ignored), methods based on generalized angular momentum representations, by analogy between two flavor oscillations and spin systems, can reach up to thousands of neutrinos and predict the thermodynamic limit~\cite{Friedland:2003eh, Friedland:2006ke, Xiong:2022, Roggero:2022}. 

In the case of time-dependent interaction strength and all-to-all $\nu$-$\nu$ interactions, the more sophisticated tensor network method, namely, the time-dependent variational principle (TDVP) method has been utilized in~\cite{Cervia:2022}. These techniques provided considerable computational benefit for an initial state with all neutrinos in the same flavor, allowing for evolution of a system with $\approx50$ oscillation modes. This was a consequence of the entanglement among neutrinos being more localized in certain regions of the neutrino energy distribution. For systems with initial states being a mixture of $\nu_e$ and $\nu_x$ flavors, the entanglement is more de-localized, and therefore, the comparative advantage gained through TDVP methods is less dramatic, although work remains in progress on this front.  

For a general setup, quantum computers are a promising tool to solve the quantum many-body problem. Initial steps~\cite{Hall:2021rbv,Yeter-Aydeniz:2021olz,Illa:2022jqb,Amitrano:2022yyn} to simulate the collective neutrino oscillations on a quantum computer are already taken in this direction. In \cite{Hall:2021rbv} a sytem of four neutrinos was simulated on IBM's quantum devices using the real-time evolution. The unitary evolution operator $U(t)=\exp(-\mathrm{i}Ht)$ was decomposed using the first order Trotter-Suzuki decomposition, where error scales as $\mathcal{O}(t^{2})$. Since the interaction is long-range, a device with all-to-all connectivity among qubits is preferred. As an alternative, SWAP operations have been used to implement this interaction on a quantum device having connectivity among neighboring qubits~\cite{Hall:2021rbv}. In \cite{Yeter-Aydeniz:2021olz}, the hybrid quantum-classical algorithm QLanczos (quantum Lanczos) was used to calculate the eigenvalues of neutrino many-body interaction Hamiltonian~\cite{Patwardhan:2019zta} on a quantum computer. Furthermore, the transition probabilities of collective neutrino oscillations were obtained by performing the real-time evolution using trotterization. However, all these earlier quantum computing studies were limited to a small system of four neutrinos due to constraints in the form of currently available quantum devices, which can perform only a limited number of operations with low accuracy. More recently in~\cite{Amitrano:2022yyn}, a trapped-ion quantum device was utilized to perform the simulations for up to eight neutrinos, thanks to the all-to-all qubit connectivity in trapped-ion based architecture. 

\section{Concluding remarks}

Studying the many-body quantum dynamics of dense neutrino systems remains an active area of research, with various groups attempting to investigate the problem using different types of classical and quantum computational tools, as well as analytic or semi-analytic descriptions. In environments where neutrinos are present in high number densities, they almost inevitably become the main carriers of energy and lepton number, and as a result, the physics of neutrino flavor transformation in these environments becomes particularly relevant for the dynamics and nucleosynthesis. Moreover, the close parallels between this problem and other quantum many-body systems in nuclear and condensed-matter physics suggests that the results and insights obtained through these studies could have a much broader scope, beyond just the field of neutrino physics.




\bibliographystyle{hapalike}
\bibliography{references}

\end{document}